\documentclass[11pt]{article}

\def\ket#1{\left| #1 \right\rangle}

\def\problem#1{{\bf #1}}
\def\half{\textstyle{1 \over 2}}
\def\s2{{\textstyle{1 \over \sqrt{2}}}}

\def\ra{\rightarrow}
\def\xor{\oplus}
\def\01{\{0,1\}}
\def\e{\varepsilon}
\def\x{\times}

\def\ni{\noindent}

\def\cents{\hbox{\sf\rlap/c}}
\def\a{\alpha}
\def\b{\beta}
\def\aa{\alpha^{\prime}}
\def\bb{\beta^{\prime}}

\newcommand{\jbox}{\hfill\rule{3.3mm}{3.3mm}}

\newtheorem{theorem}{Theorem}

\title{An Introduction to Quantum Complexity Theory}

\author{Richard Cleve \\
{\small\sl University of Calgary}%
\footnote{Department of Computer Science, University of Calgary, 
Calgary, Alberta, Canada T2N 1N4. Email: {\tt cleve@cpsc.ucalgary.ca}.}}

\date{}

\begin{document}

\maketitle

\begin{abstract}
We give a basic overview of computational complexity, query complexity, 
and communication complexity, with quantum information incorporated 
into each of these scenarios.
The aim is to provide simple but clear definitions, and to highlight 
the interplay between the three scenarios and currently-known quantum 
algorithms.
\end{abstract}

{\em Complexity theory\/} is concerned with the inherent cost required 
to solve information processing problems, where the cost is measured 
in terms of various well-defined resources.
In this context, a {\em problem\/} can usually be thought of as a 
function whose input is a {\em problem instance\/} and whose 
corresponding output is the {\em solution\/} to it.
Sometimes the solution is not unique, in which case the problem can 
be thought of as a relation, rather than a function.
{\em Resources} are usually measured in terms of: some designated 
elementary operations, memory usage, or communication.
We consider three specific complexity scenarios, which illustrate 
different advantages of working with quantum information: 
\begin{enumerate}
\item
{\bf Computational complexity} 
\item
{\bf Query complexity}
\item
{\bf Communication complexity}.
\end{enumerate}
Despite the differences between these models, there are some intimate 
relationships among them.
The usefulness of many currently-known quantum algorithms is 
ultimately best expressed in the computational complexity model; 
however, virtually all of these algorithms evolved from algorithms 
in the query complexity model.
The query complexity model is a natural setting for discovering 
interesting quantum algorithms, which frequently have interesting 
counterparts in the computational complexity model.
Quantum algorithms in the query complexity model can also be transformed 
into protocols in the communication complexity model that use quantum 
information (and sometimes these are more efficient than any classical 
protocol can be).
Also, this latter relationship, taken in its contrapositive form, 
can be used to prove that some problems are inherently difficult in 
the query complexity model.

\section{Computational complexity}

In the {\em computational complexity\/} scenario, an {\em input\/} is 
encoded as a binary string (say) and supplied to an algorithm, which 
must compute an {\em output\/} string corresponding to the input.
For example, in the case of the factoring problem, for input 
100011 (representing 35 in binary), the valid outputs might be 000101 
or 000111 (representing the factors of 35).
The algorithm must produce the required output by a series of {\em local\/} 
operations.
By this, we do not necessarily mean ``local in space'', but, 
rather, that each operation involves a small portion of the data.
In other words, a local operation is a transformation that is confined to 
a small number of bits or qubits (such as two or three).
The above property is satisfied by Turing machines and circuits, and 
also by quantum Turing machines \cite{BV93,Deu85} and quantum circuits 
\cite{Deu89,Yao93} (see also \cite{NO99}).
We shall find it most convenient to work with circuit models here.

\subsection{Classical circuits}

For classical circuits, the basic operations can be taken as 
the binary $\wedge$ ({\sc and}) gate, the binary $\vee$ ({\sc or}) gate, 
and the unary $\neg$ ({\sc not}) gate.
In Fig.~1 is a boolean circuit consisting of five gates that 
computes the parity of two bits.
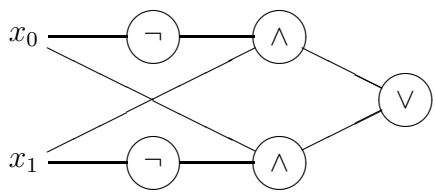
\begin{figure}[h]
\centering

\setlength{\unitlength}{0.033in}

\begin{picture}(80,30)(0,7)

\put(30,10){\circle{8}}
\put(30,30){\circle{8}}
\put(50,10){\circle{8}}
\put(50,30){\circle{8}}
\put(70,20){\circle{8}}

\put(26,6){\makebox(8,8){$\neg$}}
\put(26,26){\makebox(8,8){$\neg$}}
\put(46,6){\makebox(8,8){$\wedge$}}
\put(46,26){\makebox(8,8){$\wedge$}}
\put(66,16){\makebox(8,8){$\vee$}}
\put(9.5,30){\makebox(0,0){{\large $x$}$_0$}}
\put(9.5,10){\makebox(0,0){{\large $x$}$_1$}}

\put(13.4,10){\line(1,0){12.5}}
\put(13.4,30){\line(1,0){12.5}}

\put(34.1,10){\line(1,0){11.9}}
\put(34.1,30){\line(1,0){11.9}}

\put(13.2,11.7){\line(2,1){33.0}}
\put(13.2,28.3){\line(2,-1){33.0}}

%\put(53.4,11.9){\line(2,1){13.4}}
%\put(53.4,28.3){\line(2,-1){13.4}}
\put(53.6,12.0){\line(2,1){12.7}}
\put(53.6,28.2){\line(2,-1){12.7}}

\end{picture}

\caption{\small A classical circuit for computing the parity of two bits.}
\label{fig1}
\end{figure}
The inputs are denoted as $x_0$ and $x_1$, and the ``data-flow'' 
is from left to right.
The rightmost gate is designated as the output, whose value 
is $x_0 \xor x_1$, as required.
This is the smallest circuit consisting of $\wedge$, $\vee$, and $\neg$ 
gates that computes the parity.
Based on this fact, we could say that the computational complexity of 
the binary parity function is five.
But note that this value is highly dependent on the specific set of 
basic operations that we started with.
If we included the binary $\xor$ ({\sc exclusive-or}) gate as a 
basic operation then a single gate suffices to compute the parity of 
two bits (Fig.~\ref{fig2}).
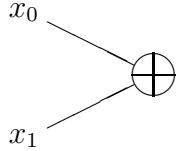
\begin{figure}[h]
\centering

\setlength{\unitlength}{0.033in}

\begin{picture}(40,30)(0,7)

\put(30,20){\circle{7}}
\put(26.5,20){\line(1,0){7}}
\put(30,16.5){\line(0,1){7}}

\put(9.5,30){\makebox(0,0){{\large $x$}$_0$}}
\put(9.5,10){\makebox(0,0){{\large $x$}$_1$}}

\put(13.2,11.6){\line(2,1){13.7}}
\put(13.2,28.4){\line(2,-1){13.7}}

\end{picture}

\caption{\small An alternative circuit for parity with one 
exclusive-or gate.}
\label{fig2}
\end{figure}

This illustrates a feature of the computational complexity model: 
the {\em exact\/} number of operations required to compute functions 
is quite sensitive to the technical choice of which basic operations 
to allow.
The exact computational complexity of simple problems involving a 
small number of bits is somewhat arbitrary.

Computational complexity is more meaningful when larger problems 
that scale up are considered, such as the problem of computing the 
parity of $n$ bits, $x_0,x_1,\ldots,x_{n-1}$.
Using $\xor$ gates, one can construct a tree with $n-1$ such gates 
that computes this parity.
On the other hand, if only $\wedge$, $\vee$, and $\neg$ gates 
are available then it appears that something like $5(n-1)$ gates 
are needed.
In both cases, the number of gates is $O(n)$, and it is also 
straightforward to prove that a constant times $n$ gates are 
{\em necessary\/} for both cases.
A similar property holds for {\em any\/} computational complexity 
problem: changing from one set of gates to any other set of gates 
(assuming that both sets are local and universal) can only affect 
computational complexity by a multiplicative constant.
Thus, for any $f : \01^{\ast} \rightarrow \01$, its computational 
complexity is a well-defined function (of $n$, the length 
of the input to $f$) up to a multiplicative constant.

This is one reason why it is common to denote the computational complexity 
of functions using asymptotic notation that suppresses multiplicative 
constants.
$O(T(n))$ means bounded above by $c\,T(n)$ for some constant $c > 0$ 
(for sufficiently large $n$).
Also, $\Omega(T(n))$ means bounded below by $c\,T(n)$ for some constant 
$c > 0$, and $\Theta(T(n))$ means both $O(T(n))$ and $\Omega(T(n))$.
A circuit is {\em polynomially-bounded\/} in size if its size 
is $O(n^d)$ for some constant $d$.

A matter that we have so far obscured concerns the treatment of the 
parameter $n$ (denoting the input size).
Although each circuit is for some {\em fixed\/} value of $n$, we are 
also speaking of $n$ as a freely varying parameter.
For problems where $n$ is a variable (such as the problem of computing 
the parity of $n$ bits), an algorithm in the circuit model must 
actually be a {\em circuit family\/} $(C_1, C_2, C_3, \ldots)$, where 
circuit $C_n$ is responsible for all input instances of size $n$.
To be meaningful, a circuit family has to be {\em uniform\/} in that 
it can somehow be finitely specified.
For example, for the aforementioned parity problem, a finite 
specification of a circuit family can be informally: ``for input size $n$, 
$C_n$ is a binary tree of $\xor$-gates with $x_0,\ldots,x_{n-1}$ 
at the leaves''.
Formally, a {\em specification\/} of a circuit family is an algorithm 
that maps each $n$ to an explicit description of $C_n$.
Technically, we ought to include the efficiency of the 
specification algorithm as part of the computational cost of 
a circuit family.
This raises the question of what formalism one uses to describe the 
specification algorithm.
Note that if we try to use another circuit family for this then 
{\em it\/} requires its own specification algorithm (and so on!), 
so this approach will not work.
There are sophisticated ways of dealing with uniformity; a very simple 
way is to just use some non-circuit model, such as a Turing machine 
(running in time, say, polynomial in $n$) for the circuit specification 
algorithm.
At this point, the reader may wonder why one does not just use 
the Turing machine model to begin with.
A big advantage of circuits is that their structural elements are 
simple and easy to work with---and this appears to hold for quantum 
circuits as well.
Uniformity tends to be a straightforward technicality, that can be worked 
out after a circuit family is discovered; the discovery of the circuit 
family is usually the interesting part of the algorithm design process.

Let us now consider the problem of 
\problem{primality testing}, 
where the input is a number $x$ represented as an $n$-bit binary string, 
and the output is (say) 1 if $x$ is prime and 0 if $x$ is composite.
Notice that, in the cases where $x$ is composite, there is no requirement 
here that a factor of $x$ be produced.
It turns out that the smallest currently-known uniform circuit family 
for this problem has size  $O(n^{d \log \log n})$ (for some constant $d$), 
which is shy of being polynomially-bounded \cite{APR83}.

There exist {\em probabilistic\/} circuit families that solve primality 
testing more efficiently.
A {\em probabilistic circuit\/} is one that can flip coins during 
its execution, and the evolution of the computation can depend on the 
outcomes.
Formally, a $\cents$ ({\sc coin-flip}) gate, has no input and 
is understood to emit one uniformly-distributed random bit when 
executed during a computation.
If $m$ random bits are required then $m$ $\cents$-gates can be 
inserted into a circuit.
Solovay and Strassen \cite{SS77} discovered a remarkable probabilistic 
algorithm for primality testing that can be expressed in terms of 
probabilistic circuits.
For any $\e > 0$, there is a probabilistic circuit of size 
$O(n^3\log(1 / \e))$ that errs with probability at most $\e$.
That is, given any $x \in \01^n$ as input, the circuit correctly 
decides the primality of $x$ with probability at least $1 - \e$. 
Note that the error probability is with respect to the $\cents$-gates, 
and not with respect to any assumed probability distribution on the 
input $x$.
The circuit family is highly uniform, and there are versions of the 
algorithm that are quite efficient in practice, even when $\e$ is very 
small (such as one billionth).

As an aside, we note that probabilistic circuit families can be 
translated into standard (deterministic) circuit families if one is 
willing to forfeit uniformity.
For each $n$, by setting $\e = 1 / (2^n+1)$, we obtain a probabilistic 
circuit $C_n$ of size $O(n^4)$ for primality testing that errs with 
probability less than $1 / 2^n$ for any input.
Now consider the circuit $C^{\prime}_n$ that results if, for each 
$\cents$-gate in $C_n$, a uniformly distributed random bit is independently 
generated and substituted for that gate.
This is a probabilistic construction that yields a deterministic 
circuit $C^{\prime}_n$.
For $x \in \01^n$, let $p_x$ be the probability that the resulting 
$C^{\prime}_n$ errs on input $x$.
Then, for each $x$, $p_x < 1 / 2^n$, so the probability that $C^{\prime}_n$ 
errs for {\em any\/} $x \in \01^n$ is strictly less than 
$\sum_{x \in \01^n} 1 / 2^n = 1$.
Therefore, with probability greater than 0, $C^{\prime}_n$ is correct 
for all of its $2^n$ possible input values.
It follows that, for any $n$, a deterministic circuit of size $O(n^4)$ 
must exist for primality testing.
The problem is that there is no known efficient way to explicitly 
construct the coin flips which yield a correct circuit.
Thus, the implied $O(n^4)$-size circuit family for primality testing 
is merely established by an existence proof; this is an example of a 
{\em non-uniform\/} circuit family.
The fact that uniform probabilistic circuit families can be converted 
into non-uniform deterministic circuit families is theoretically noteworthy, 
but not practical.

A problem that is related to---but different from---primality testing 
is the 
\problem{factoring problem}, 
where the input 
is an $n$-bit number $x$, and the output is a list of the prime 
factors of $x$.
This is apparently much harder than primality testing, since 
the smallest currently-known circuit family for this problem 
is probabilistic and has size $O(2^{^{d \sqrt{n \log n}}})$ (where 
$d$ is a constant) \cite{LP92,Pom87}, which is far from being 
polynomially-bounded.
One of the reasons why quantum algorithms are of interest is that 
there exists a {\em quantum\/} circuit family of polynomial-size 
that solves the factoring problem (this will be discussed later).

A problem that is closely related to the factoring problem is the 
\problem{order-finding problem}, 
where the input is a pair of natural numbers $a$ and $N$ that are 
coprime (i.e.\ such that $\gcd(a,N) = 1$), and the goal is to find 
the smallest positive $r$ such that $a^r \bmod N = 1$ (there 
always exists such an $r \in \{1,\ldots,N-1\}$).
It turns out that a polynomial-size circuit family for the 
order-finding problem can be converted into a polynomial-size 
probabilistic circuit family for the factoring problem 
(and vice versa).
In fact, the quantum circuit for factoring is actually obtained via 
this relationship from a quantum circuit that solves the order-finding 
problem.

Although we have represented circuits pictorially as data-flow diagrams, 
it is useful to be able to encode circuits as binary strings.
There are several reasonable encoding schemes.
One such scheme encodes the graphical structure of a circuit $C$ as 
an $m \times m$ adjacency matrix (where $m$ is the number of gates plus 
the number of inputs in $C$), and then follows this by more bits 
that specify the labels of the nodes (e.g.\ $\wedge$, $\vee$, $\neg$, 
$x_0,\ldots,x_{n-1}$).
Note that, using this encoding scheme, a circuit of size $m$ has an encoding 
of $O(m^2)$ bits.
There are more efficient encoding schemes, where the encodings are of 
length $O(m \log m)$, and, for any ``reasonable'' encoding scheme, 
the length of the string that encodes $C$ is polynomially-related to 
the size of $C$.
Let $e(C)$ denote a binary string that encodes the circuit $C$ (relative 
to some reasonable encoding scheme).

A fundamental problem in classical computational complexity 
is the \problem{circuit satisfiability problem}, which is defined as 
follows.
Call a circuit {\em satisfiable\/} if there exists an input string 
to the circuit for which the corresponding output value of the 
circuit is 1.
For example, the circuit in Fig.~\ref{fig1} is satisfiable.
The input to the circuit satisfiability problem is a binary string 
$x = e(C)$ that encodes some boolean circuit $C$, and the output 
is 1 if $C$ is satisfiable, and 0 otherwise.
The best currently-known (deterministic or probabilistic) algorithm 
for circuit satisfiability is to simply try all possible inputs to $C$.
When $e(C)$ encodes a circuit $C$ with $n$ inputs and $m$ gates, 
this procedure takes $O(2^{n}m^d)$ steps, where $d$ is a constant that 
depends on the encoding scheme used ($d=2$ suffices for most reasonable 
encoding schemes).
In interesting cases, $m$ is typically polynomial in $n$, so the dominant 
factor in this quantity is $2^n$.
It is not known whether or not there is a polynomially-bounded 
circuit family for circuit satisfiability.
In fact, circuit satisfiability is one of the so-called 
${\it NP\/}$-complete problems \cite{Coo71,GJ79}, for which a 
polynomially-bounded circuit family would yield polynomially-bounded 
circuits for {\em all\/} problems in ${\it NP}$.

\subsection{Quantum circuits}

To develop a theory of computational complexity for quantum information, 
it is natural to extend the notion of a circuit to a composition gates 
which perform {\em quantum\/} operations on {\em quantum bits\/} 
(called {\em qubits}).
The most general quantum operations subsume all classical operations, 
which are frequently not reversible.
It turns out that the quantum operations that seem to be the most 
useful in the context of quantum computation are those that 
are unitary (and hence reversible), as well as the von Neumann 
measurements.

Let us begin by recalling that the state of a system of $m$ qubits 
can be described by associating an {\em amplitide} $\a_x$ with each 
$x \in \01^m$ (we restrict our attention to {\em pure\/} quantum states).
Each amplitude is a complex number and there is a condition that 
$\sum_{x \in \01^m} |\a_x|^2 = 1$.
Taken together, these amplitudes constitute a point in a $2^m$-dimensional 
vector space.
The {\em computational basis\/} associated with this vector space 
is $\{ \ket{x} : \mbox{$x \in \01^m$} \}$, and we follow the convention 
of writing states as linear combinations of these basis elements: 
\begin{eqnarray}
\sum_{x \in \01^m} \a_x \ket{x}.
\end{eqnarray}
%The vector space is endowed with the unique inner product with respect 
%to which the computational basis is orthonormal.
Given a quantum state, it is impossible to access the values of the 
amplitudes directly.
What one can do is perform a (von Neumann) measurement on each qubit.
If such an operation is performed then the result is a random $m$-bit 
string $y$, distributed as $\Pr[y = x] = |\a_x|^2$, for each 
$x \in \01^m$.
After this measurement, the original quantum state is destroyed.
One can also perform a unitary operation on an $m$-qubit system, 
which is a linear transformation $U$ for which 
$U U^{\mbox{\scriptsize \dag}} = I$, where $U^{\mbox{\scriptsize \dag}}$ 
is the conjugate transpose of $U$.
Such a unitary transformation can be represented by a $2^m \times 2^m$ 
matrix and will, in general, affect all of the $m$ qubits.

For the purposes of quantum computation, we restrict the basic operations 
to local unitary transformations that only involve a small number 
(say, one or two) of the qubits.
A one-qubit unitary operation can be described by $2 \x 2$ unitary 
matrix $U$.
In the case where $m=1$, this $U$ transforms the state 
$\a \ket{0} + \b \ket{1}$ to the state $\aa \ket{0} + \bb \ket{1}$, 
where 
\begin{eqnarray}
\pmatrix{\aa \cr \bb} & = & U \pmatrix{\a \cr \b}.
\end{eqnarray}

In order to define the semantics of applying a one-qubit gate in 
the context of an $m$-qubit system for $m > 1$, we introduce 
a tensor product operation.
Suppose that an $m$-qubit system is in state 
$\sum_{x \in \01^m} \a_x \ket{x}$ and an $n$-qubit system is in 
state $\sum_{y \in \01^n} \b_y \ket{y}$.
Then the state of the combined system (consisting of $m+n$ qubits) 
is defined to be the {\em tensor product\/} of the states of the 
individual systems, which is 
\begin{eqnarray}
\left(\sum_{x \in \01^m} \a_x \ket{x}\right) 
\left(\sum_{y \in \01^n} \b_y \ket{y}\right) 
& = & 
\sum_{x \in \01^m \atop y \in \01^n} \a_x\b_y \ket{xy}.
\end{eqnarray}
For example, 
$(\s2\ket{0}-\s2\ket{1}) (\s2\ket{0}-\s2\ket{1}) 
= \half\ket{00} - \half\ket{01} - \half\ket{10} + \half\ket{11}$.
Now, applying a one-qubit $U$ to the $k^{\mbox{\scriptsize th}}$ 
qubit of an $m$-qubit system is defined to be the unitary 
operation that maps each basis state 
$$\ket{x_0 \cdots x_{m-1}} = 
\ket{x_0 \cdots x_{k-2}}\ket{x_{k-1}}\ket{x_k \cdots x_{m-1}}$$ 
to the state 
$$\ket{x_0 \cdots x_{k-2}}(U\ket{x_{k-1}})\ket{x_k \cdots x_{m-1}}$$ 
(for each $x \in \01^m$).
Note that, by linearity, this completely defines a unitary operation 
on an $m$-qubit system.

\begin{figure}[h]
\centering

\setlength{\unitlength}{0.033in}

\begin{picture}(30,30)(0,10)

\put(0,15){\line(1,0){10}}
\put(20,15){\line(1,0){10}}
\put(0,30){\line(1,0){30}}

\put(10,10){\framebox(10,10){$H$}}

\end{picture}

\caption{\small Quantum circuit applying a Hadamard gate to one 
of two qubits.}
\label{fig3}
\end{figure}
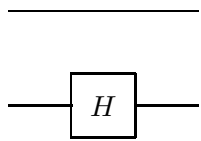

For example, the one-qubit {\em Hadamard\/} gate corresponds to the 
matrix 
\begin{eqnarray}
\label{hadamard}
H & = & \s2 \pmatrix{1 & \ \ 1 \cr 1 & -1},
\end{eqnarray}
and, when it is applied to the second qubit of a two-qubit system, 
the resulting operation is 
\begin{eqnarray}
\s2\pmatrix{1 & \ \ 1 & \ \ 0 & \ \ 0 \cr
            1 &    -1 & \ \ 0 & \ \ 0 \cr
            0 & \ \ 0 & \ \ 1 & \ \ 1 \cr
            0 & \ \ 0 & \ \ 1 &    -1}
\end{eqnarray}
(with respect to the ordering of basis states $\ket{00}$, $\ket{01}$, 
$\ket{10}$, $\ket{11}$).
%in state $\s2\ket{00} + \s2\ket{11}$ then the state changes to 
%$\half\ket{00} + \half\ket{01} + \half\ket{10} - \half\ket{11}$.
A quantum circuit corresponding to such an operation is in 
Fig.~\ref{fig3}, which denotes that the first (top) qubit is left 
unaltered and $H$ is applied to the second qubit.

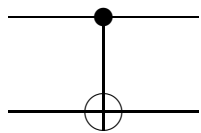
\begin{figure}[h]
\centering

\setlength{\unitlength}{0.033in}

\begin{picture}(30,30)(0,10)

\put(0,15){\line(1,0){30}}
\put(0,30){\line(1,0){30}}

\put(15,30){\circle*{3}}

\put(15,12){\line(0,1){18}}

\put(15,15){\circle{6}}

\end{picture}

\caption{\small Notation for the {\sc controlled-not} ({\sc c-not}) gate.}
\label{fig4}
\end{figure}

To construct nontrivial quantum circuits, it is necessary to 
include two-qubit unitary operations.
A simple but quite useful two-qubit operation is the 
{\sc controlled-not} gate ({\sc c-not}, for short), which, for all 
$x, y \in \01$, transforms the basis state $\ket{x}\ket{y}$ to the 
basis state $\ket{x}\ket{y \xor x}$ (and this extends to arbitrary 
quantum states by linearity).
The notation for the {\sc c-not} gate in quantum circuits is 
indicated in Fig.~\ref{fig4} (it is also known as the ``reversible 
exclusive-or'' gate).

\ni Note that the {\sc c-not} gate corresponds to the unitary transformation 
\begin{eqnarray}
\pmatrix{1 & 0 & 0 & 0 \cr 
         0 & 1 & 0 & 0 \cr 
         0 & 0 & 0 & 1 \cr 
         0 & 0 & 1 & 0}.
\end{eqnarray}
The semantics of the {\sc c-not} gate extends to the context 
of $m$-qubit systems with $m > 2$ in a manner similar to that of 
the one-qubit gates.

For basis states $\ket{x}\ket{y}$, the effect of the {\sc c-not} gate 
is essentially the same as the classical two-bit gate that maps 
$(x,y)$ to $(x,x \xor y)$ (for all $x, y \in \01$).
This gate negates the second bit conditional on the first bit being 1.
For arbitrary quantum states, the behavior of this gate is more subtle.
For example, although the classical gate never changes the value of 
its first ``control'' bit, the quantum gate sometimes does: applying 
the {\sc c-not} gate to state 
$(\s2\ket{0}-\s2\ket{1})(\s2\ket{0}-\s2\ket{1})$ 
yields the state 
$(\s2\ket{0}+\s2\ket{1})(\s2\ket{0}-\s2\ket{1})$.

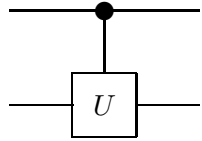
\begin{figure}[h]
\centering

\setlength{\unitlength}{0.033in}

\begin{picture}(30,30)(0,10)

\put(0,15){\line(1,0){10}}
\put(20,15){\line(1,0){10}}
\put(0,30){\line(1,0){30}}

\put(15,30){\circle*{3}}

\put(15,20){\line(0,1){10}}

\put(10,10){\framebox(10,10){$U$}}

\end{picture}

\caption{\small Notation for a {\sc controlled}-$U$ gate.}
\label{fig5}
\end{figure}

A more general kind of two-qubit gate is the {\sc controlled}-$U$ 
gate, where $U$ is a $2 \times 2$ unitary matrix.
This gate maps $\ket{0}\ket{y}$ to $\ket{0}\ket{y}$ and 
$\ket{1}\ket{y}$ to $\ket{1}(U\ket{y})$ (for all $y \in \01$), 
and is denoted in Fig.~\ref{fig5}.

\ni Note that the {\sc c-not} gate is a special case of a 
{\sc controlled}-$U$ gate with 
\begin{eqnarray}
U = \pmatrix{0 & 1 \cr 1 & 0}
\end{eqnarray}
(and this $U$ itself is essentially a {\sc not} gate).

\begin{figure}[h]
\centering

\setlength{\unitlength}{0.0308in}

\begin{picture}(120,45)(0,10)

\put(0,15){\line(1,0){10}}
\put(20,15){\line(1,0){5}}
\put(35,15){\line(1,0){20}}
\put(65,15){\line(1,0){20}}
\put(95,15){\line(1,0){5}}
\put(110,15){\line(1,0){10}}

\put(0,30){\line(1,0){120}}
\put(0,45){\line(1,0){120}}

\put(45,30){\circle{6}}
\put(75,30){\circle{6}}

\put(30,30){\circle*{3}}
\put(60,30){\circle*{3}}
\put(45,45){\circle*{3}}
\put(75,45){\circle*{3}}
\put(90,45){\circle*{3}}

\put(90,20){\line(0,1){25}}
\put(30,20){\line(0,1){10}}
\put(45,27){\line(0,1){18}}
\put(60,20){\line(0,1){10}}
\put(75,27){\line(0,1){18}}

\put(10,10){\framebox(10,10){$H$}}
\put(25,10){\framebox(10,10){$V$}}
\put(55,10){\framebox(10,10){$V^{\mbox{\scriptsize \dag}}$}}
\put(85,10){\framebox(10,10){$V$}}
\put(100,10){\framebox(10,10){$H$}}

\end{picture}
\begin{picture}(14,45)(0,2.5)

\put(0,0){\makebox(14,45){$\equiv$}}

\end{picture}
\begin{picture}(22,45)(0,10)

\put(0,15){\line(1,0){22}}
\put(0,30){\line(1,0){22}}
\put(0,45){\line(1,0){22}}

\put(11,30){\circle*{3}}
\put(11,45){\circle*{3}}

\put(11,12){\line(0,1){33}}

\put(11,15){\circle{6}}

\end{picture}

\caption{\small Quantum circuit simulating a {\sc c$^2$-not} 
(Toffoli) gate.}
\label{fig6}
\end{figure}
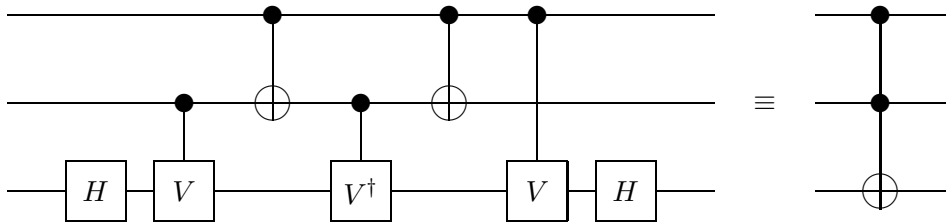

Now, suppose that we want to compute the {\sc and} of two bits 
(i.e.\ take $x_0$ and $x_1$ as input and produce $x_0 \wedge x_1$ 
as output) using only the one- and two-qubit gates of the above form.
This can be done in a manner that avoids irreversible operations 
via the quantum circuit in Fig.~\ref{fig6}, where $H$ is the Hadamard 
gate (Eq.~\ref{hadamard}) and 
\begin{eqnarray}
\label{vmatrix}
V & = & \pmatrix{1 & 0 \cr 0 & i}
% \ \ \ \mbox{(where $i = \sqrt{-1}$).}
\end{eqnarray}
(where $i = \sqrt{-1}$).

\ni For any $x_0, x_1, y \in \01$, setting the initial state of the 
qubits to $\ket{x_0}\ket{x_1}\ket{y}$ and tracing through the execution 
of this circuit reveals that the final state 
is $\ket{x_0}\ket{x_1}\ket{y \xor (x_0 \wedge x_1)}$.
Thus, when $y$ is initialized to 0, the final state of the 
third qubit is $\ket{x_0 \wedge x_1}$ (and the explicit classical 
data, $x_0 \wedge x_1$, can be extracted from this quantum state by 
a measurement).
The three-qubit operation that is simulated in Fig.~\ref{fig6} is a 
so-called Toffoli gate (also called a {\sc controlled-controlled-not}, 
or {\sc c$^2$-not} for short).
See \cite{BBC95,Div98,SW95} for some similar constructions.

For classical circuits, there are finite sets of gates which are 
universal in the sense that they can be used to simulate any other 
set of gates.
For quantum circuits, the situation is different, since the 
set of all unitary operations is continuous, and hence 
uncountable---even when restricted to one-qubit gates.
If one starts with any finite set of quantum gates then the set 
of all unitary operations that can implemented is limited to some 
countable subset of all the unitary operations.
In spite of this, there are meaningful ways to capture the 
important features associated with universal sets of gates.

First, it turns out that there are {\em infinite\/} sets 
consisting of one- and two-qubit of gates that are universal 
in the exact sense.
For example, if the {\sc c-not} gate as well as all unitary 
one-qubit gates are available then any $k$-qubit unitary 
operation can be simulated with $O(4^k k)$ such gates 
\cite{BBC95,Kni95}.
Therefore, the overhead is constant when switching between 
different universal sets of local unitary gates (such as the 
set of all two-qubit gates and the set of all three-qubit gates).

Moreover, there are {\em finite\/} sets of one- and two-qubit 
gates that are universal in an {\em approximate\/} sense.
The aforementioned one-qubit Hadamard gate $H$ (Eq.~\ref{hadamard}) 
and the two-qubit {\sc controlled}-$V$ gate (where $V$ is defined 
in Eq.~\ref{vmatrix}) are an example of such a set.
The precise result is best stated as a theorem.

\begin{theorem}[\cite{Kit97,Sol99}]
\label{approximate}
Let $B$ be any two-qubit gate and $\e > 0$.
Then there exists a quantum circuit of size $O(\log^d(1 / \e))$ 
(where $d$ is a constant) consisting of only $H$ and {\sc controlled}-$V$ 
gates which computes a unitary operation $B^{\prime}$ that approximates 
$B$ in the following sense.
There exists a unit complex number $\lambda$ (i.e.\ with $|\lambda| = 1$) 
such that $||B - \lambda B^{\prime}||_2 \le \e$.
\end{theorem}

\ni In the above theorem, $||\cdot||_2$ is the norm induced by Euclidean 
distance and $\lambda$ is a ``global phase factor''(which can be 
disregarded).
Consequently, if $B^{\prime}$ is substituted for $B$ in some quantum 
circuit then the final state $\sum_{x} \aa_x \ket{x}$ approximates the 
final state of the original circuit $\sum_{x} \a_x \ket{x}$ in the sense 
that $\sqrt{\sum_{x} |\lambda \aa_x - \a_x|^2} \le \e$.
This implies that if the final state is measured then the 
probability of any event among the possible outcomes is affected 
by at most $\e$.
The proof of Theorem~\ref{approximate} exploits the fact that the 
commutator of two unitary operators is not generally $I$ (the 
identity operator), but it can converge very quickly to $I$ 
(see \cite{Kit97,Sol99} for details).

\begin{figure}[h]
\centering

\setlength{\unitlength}{0.0308in}

\begin{picture}(70,30)(0,10)

\put(0,15){\line(1,0){10}}
\put(20,15){\line(1,0){20}}
\put(50,15){\line(1,0){20}}

\put(0,30){\line(1,0){10}}
\put(20,30){\line(1,0){50}}

\put(30,15){\circle{6}}
\put(60,15){\circle{6}}

\put(30,30){\circle*{3}}
\put(60,30){\circle*{3}}

\put(30,12){\line(0,1){18}}
\put(60,12){\line(0,1){18}}

\put(10,10){\framebox(10,10){$W$}}
\put(10,25){\framebox(10,10){$W$}}
\put(40,10){\framebox(10,10){$W^{\mbox{\scriptsize \dag}}$}}

\end{picture}
\begin{picture}(14,30)(0,3.0)

\put(0,0){\makebox(14,30){$\equiv$}}

\end{picture}
\begin{picture}(30,30)(0,10)

\put(0,15){\line(1,0){10}}
\put(20,15){\line(1,0){10}}
\put(0,30){\line(1,0){30}}

\put(15,30){\circle*{3}}

\put(15,20){\line(0,1){10}}

\put(10,10){\framebox(10,10){$V$}}

\end{picture}

\caption{\small Simulation of a {\sc controlled-$V$} gate 
(note: $W^{\mbox{\scriptsize \dag}} = W^7$).}
\label{fig6.1}
\end{figure}
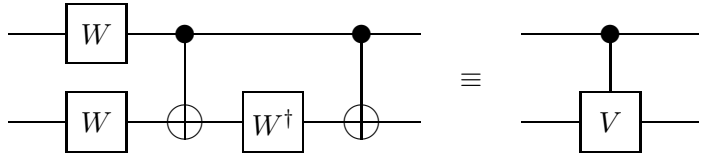

An example of another finite set of gates that is universal in the 
approximate sense is: $H$, $W$, and {\sc c-not}, where 
\begin{eqnarray}
\label{wmatrix}
W & = & \pmatrix{1 & 0 \cr 0 & \ e^{i \pi / 4}}.
\end{eqnarray}
In fact, with $W$ and {\sc c-not} gates, one can simulate a 
{\sc controlled-$V$} gate, as shown in Fig.~\ref{fig6.1} 
(see also \cite{BMPRV99}).

As in the classical case, the measure of computational complexity 
for quantum circuits is most interesting when large problems that scale 
up are considered.
Using sets of gates that are universal in the exact sense, computational 
complexity can vary only by constant factors.
On the other hand, using sets of gates that are universal in the 
approximate sense, computational complexity can vary by at most 
polylogarithmic factors: any circuit with $m$ gates can be simulated 
within accuracy $\e$ by a circuit in terms of a different set of basic 
operations with $O(m \log^d (m/ \e))$ gates.
This is accomplished by simulating each of the $m$ gates of the original 
circuit within accuracy $\e / m$, which results in a total accumulated 
error bounded by $\e$.

For example, consider the problem where the input is 
$x_0,x_1,\ldots,x_{n-1}$ and the goal is to compute 
the conjunction $x_0 \wedge x_1 \wedge \cdots \wedge x_{n-1}$.
In terms of $H$, $W$, and {\sc c-not} gates, the computational 
complexity can be shown to be $\Theta(n)$, and, with another set 
of approximately uniform gates, the complexity may be different, 
but it will remain between $\Omega(n)$ and $O(n\log^d(n / \e))$ 
(where $d$ is some constant and $\e$ is the accuracy level required).

Since it seems inconceivable that it would ever be possible to 
physically implement quantum gates with perfect accuracy, the need to 
ultimately work with approximations of quantum gates is inevitable.
Fortunately, due the unitarity of quantum operations, inaccuracies 
only scale up linearly with the number of gates involved in a circuit. 
And, if one employs quantum error-correcting codes and fault-tolerant 
techniques then even gates with constant inaccuracies (and that are 
subject to ``decoherence'') can in principle be employed in arbitrarily 
large quantum circuits \cite{AB96,KLZ96,Sho97} (see \cite{Pre98} for 
an extensive review).

For quantum circuit families, we must also consider the issue of 
uniformity: a legitimate quantum circuit family should be finitely 
specifiable in a computationally efficient way.
This can be defined as a straightforward extension of the uniformity 
definitions for classical circuit families, where the specification 
algorithm is {\em classical\/} and a finite set of gates that is 
approximately universal (such as $H$, $W$, and {\sc c-not}) is used.
All quantum algorithms proposed to date can be expressed as circuit 
families that are uniform in this sense (see \cite{NO99} for further 
comments).

Notwithstanding the above issues, a convenient practice is to allow 
perfect universal sets of gates, bearing in mind that: (a) they can always 
be approximated using any finite set of gates that is approximately 
universal with only a polylogarithmic penalty in the circuit 
size (even if the implementations of these gates are approximate); and 
(b) uniformity tends to be a straightforward technicality (at least with 
the quantum algorithms discovered so far).

Perhaps the most remarkable quantum algorithm that has been discovered 
to date is the factoring algorithm, due to Peter Shor \cite{Sho94}.

\begin{theorem}[\cite{Sho94}] 
There exists a quantum circuit family of size $O(n^2\log^d(n / \e))$ 
that solves the factoring problem within accuracy $\e$ (for some 
constant $d$).
\end{theorem}

\ni Note that this circuit size is essentially exponentially smaller 
than the most efficient known classical probabilistic circuit for 
factoring (whose size is $O(2^{^{d \sqrt{n \log n}}})$).
The quantum factoring algorithm actually follows from an algorithm 
for the order-finding problem, which in turn evolved from an algorithm 
in the query complexity model (explained in the next section).

The above result shows that, based on our current state of knowledge, 
quantum algorithms may be exponentially more efficient than classical 
algorithms for some problems.
The next result shows that the gain in computational efficiency 
cannot exceed one exponential.

\begin{theorem}
\label{quant_class}
For any $S(n)$-qubit quantum circuit with $T(n)$ gates there is a 
classical probabilistic circuit with $O(2^{S(n)} T(n)^3 \log^2(1 / \e))$ 
gates%
\footnote{The $T(n)^3 \log^2(1 / \e)$ factor can be replaced by a 
smaller but more complicated expression.}
that simulates it within accuracy $\e$ in the following sense.
After measuring the final state of the quantum circuit, the probability 
of any event among the outcomes differs from that of the classical 
circuit by at most $\e$.
\end{theorem}

\ni The idea behind the proof of Theorem~\ref{quant_class} is to store 
the values of all $2^{S(n)}$ amplitudes associated with an $S(n)$-qubit 
quantum system in classical bits (to an appropriate level of precision).
Then, for each of the $T(n)$ gates, these amplitudes are updated to 
reflect the effect of the gate.
At the end, the absolute value of the square of each amplitude is 
computed and the resulting probability distribution is sampled by 
using $\cents$-gates.
To obtain the upper bound in Theorem~\ref{quant_class}, it suffices 
to store each amplitude with $T(n) + \log(1 / \e)$ bits of precision, 
which requires $O(2^{S(n)} (T(n) + \log(1 / \e)))$ bits in all.
Since the effect of each quantum gate corresponds to multiplying the 
amplitude vector by a {\em sparse\/} $2^{S(n)} \x 2^{S(n)}$ matrix, 
this entails $O(2^{S(n)})$ arithmetic operations, which can be 
implemented by $O(2^{S(n)} (T(n) + \log(1 / \e))^2)$ bit operations.
Thus, the total number of classical gates is 
$O(2^{S(n)} T(n)(T(n) + \log(1 / \e))^2) \subseteq 
O(2^{S(n)} T(n)^3 \log^2(1 / \e))$.
Also, the measurement process can be simulated with 
$O(2^{S(n)} T(n)^2 \log^2(1 / \e))$ classical gates.

A more refined argument than the one above can be used to show that 
an $S(n)$-qubit circuit with $T(n)$ gates can be simulated using 
{\em space} that is polynomial in $S(n)$ and $T(n)$ (but still with 
an exponential number of operations), and there are also more 
esoteric computational models that subsume the power of quantum 
circuit families \cite{FR98}.

Regarding the circuit satisfiability problem, it is currently unknown 
whether or not there exists a polynomially-bounded quantum circuit 
family that solves it.
What is known is that quantum algorithms can solve this problem 
quadratically faster than the best currently-known classical 
algorithms for this problem.

\begin{theorem}
\label{sat}
There exists a quantum circuit family of size 
$O(\sqrt{2^n\log(1 / \e)}m^d)$ 
that solves the circuit satisfiability problem within accuracy $\e$ 
(for some constant $d$).
Here, $n$ and $m$ measure the size of the input instance: 
$n$ is the number of inputs to circuit $C$ and $m$ is the number 
of gates of $C$.
\end{theorem}

Note how this compares with the best currently-known classical 
circuit family for the circuit satisfiability problem, which has 
size $O(2^n m^d)$.
Both quantities are exponential, but $\sqrt{2^n}$ is nevertheless 
considerably smaller than $2^n$ for large values of $n$.
The quantum algorithm is a consequence of a remarkable algorithm 
in the query complexity model that was discovered by Lov Grover 
\cite{Gro96} (explained in the next section).

%It is also unknown whether or not this quantum algorithm is 
%optimal (for example, it is conceivable that 
%${\it P} = {\it NP}$ and there is a polynomial-size 
%{em classical\/} circuit family for this problem!).

\section{Query complexity}

This is an abstract scenario which can be thought of as a game, like 
``twenty questions''.
The goal is to determine some information by asking as few questions 
as possible.
This differs from the computational complexity scenario in that 
the ``input'' is not presented as a binary string at the beginning 
of the computation.
Rather, the input can be thought of as a ``black box'' computing 
a {\em function\/} $f : S \rightarrow T$, and the basic operations 
are {\em queries}, in which the algorithm specifies a $t$ from the 
domain of the function and receives the value $f(t)$ in response.
%(In classical complexity theory, this scenario is sometimes referred to 
%as ``the decision-tree model'' or ``learning with membership queries''.)

A natural example is that of ``polynomial interpolation'', where 
$f$ is an arbitrary polynomial of degree $d$ 
\begin{eqnarray}
f(t) & = & c_0 + c_1 t + \cdots + c_d t^d
\end{eqnarray}
%(over some finite field of size greater than $d$) 
and the 
goal is to determine the coefficients $c_0,c_1,\ldots,c_d$.
It is well known that $d+1$ queries to $f$ are necessary and 
sufficient to accomplish this.

In the classical case, an algorithm in this model can be represented 
by a circuit consisting of gates from some standard universal set 
(e.g.\ $\wedge$, $\vee$, $\neg$) plus additional gates to perform queries.
For $f : S \rightarrow T$, an {\em $f$-query\/} gate takes $t \in S$ as 
input and produces $f(t)$ as output.
In this scenario, since there are no input bits related to the problem 
instance (the problem instance is embodied in $f$), 
the inputs to the circuit are all set to constant values (such as 0).

In order to be able to adapt this model to settings involving quantum 
information, we slightly modify the form of the query gates so that they 
are reversible.
For example, for $f : \01^n \rightarrow \01$, 
define a {\em reversible $f$-query\/} gate as the mapping 
$\widetilde{f} : \01^n \x \01 \ra \01^n \x \01$ such that 
$\widetilde{f}(x,y) = (x, y \xor f(x))$ (for $x \in \01^n$ and 
$y \in \01$).
Note that, for classical algorithms, reversible $f$-queries yield 
exactly the same information as the non-reversible kind.
Any circuit that makes reversible $f$-queries can be 
converted into one that makes exactly the same number of 
non-reversible $f$-queries (and vice versa).
Henceforth, all queries will be assumed to be in reversible form.

\begin{figure}[h]
\centering

\setlength{\unitlength}{0.033in}

\begin{picture}(30,45)(0,10)

\put(0,15){\line(1,0){30}}
\put(0,30){\line(1,0){10}}
\put(20,30){\line(1,0){10}}
\put(0,45){\line(1,0){10}}
\put(20,45){\line(1,0){10}}

\put(15,15){\circle{6}}

\put(15,12){\line(0,1){13}}

\put(10,25){\framebox(10,25){\large $f$}}

\end{picture}

\caption{\small Notation for an $f$-query gate when 
$f : \01^2 \rightarrow \01$.}
\label{fig7}
\end{figure}
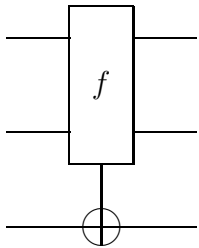

In the quantum case, an $f$-query is a unitary transformation 
that permutes the basis states according to the classical 
mapping determined by $f$ (in reversible form).
For example, for $f : \01^n \rightarrow \01$, an $f$-query gate 
is the unitary transformation that maps $\ket{x}\ket{y}$ to 
$\ket{x}\ket{y \xor f(x)}$ (for all $x \in \01^n$ and $y \in \01$).
One way of denoting $f$-queries in both classical and quantum 
circuits is shown in Fig.~\ref{fig7} (for the case where 
$f : \01^2 \rightarrow \01$).

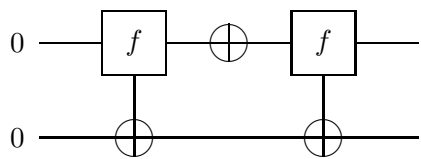
\begin{figure}[h]
\centering

\setlength{\unitlength}{0.033in}

\begin{picture}(4,30)(0,10)
\put(2,15){\makebox(0,0){$0$}}
\put(2,30){\makebox(0,0){$0$}}
\end{picture}
\begin{picture}(60,30)(0,10)

\put(0,15){\line(1,0){60}}
\put(0,30){\line(1,0){10}}
\put(20,30){\line(1,0){20}}
\put(50,30){\line(1,0){10}}

\put(15,15){\circle{6}}
\put(45,15){\circle{6}}
\put(30,30){\circle{6}}

\put(30,27){\line(0,1){6}}
\put(15,12){\line(0,1){13}}
\put(45,12){\line(0,1){13}}

\put(10,25){\framebox(10,10){$f$}}
\put(40,25){\framebox(10,10){$f$}}

\end{picture}

\caption{\small Classical circuit for Deutsch's problem using two queries.}
\label{fig8}
\end{figure}

The first instance where a quantum algorithm was proven to outperform 
any classical algorithm was with 
\problem{Deutsch's problem} \cite{Deu85}, 
where $f : \01 \ra \01$ and $f(t) = (c_0 + c_1 t) \bmod 2$, 
and the goal is to determine the value of $c_1$ (note that 
$c_1 = f(0) \xor f(1)$).
A classical circuit (in reversible form) that computes $c_1$ with 
two $f$-queries is shown in Fig.~\ref{fig8}.

\ni The inputs to the circuit are both initialized to 0, and the unary 
$\xor$ operation between the two $f$-queries is a {\sc not} gate.
%(which 
%can be regarded as an ``{\sc uncontrolled-not}'' gate).
It is easy to see that the final values of the two bits are 1 and $c_1$.
It can also be shown that no classical algorithm exists that computes 
$c_1$ with a single $f$-query (since it is impossible to determine 
$f(0) \xor f(1)$ from just $f(0)$ or $f(1)$ alone).

\begin{figure}[h]
\centering

\setlength{\unitlength}{0.033in}

\begin{picture}(6,30)(0,10)
\put(3,30){\makebox(0,0){$\ket{0}$}}
\put(3,15){\makebox(0,0){$\ket{1}$}}
\end{picture}
\begin{picture}(60,30)(0,10)

\put(0,15){\line(1,0){10}}
\put(20,15){\line(1,0){20}}
\put(50,15){\line(1,0){10}}

\put(0,30){\line(1,0){10}}
\put(20,30){\line(1,0){5}}
\put(35,30){\line(1,0){5}}
\put(50,30){\line(1,0){10}}

\put(30,12){\line(0,1){13}}

\put(30,15){\circle{6}}

\put(25,25){\framebox(10,10){$f$}}
\put(10,10){\framebox(10,10){$H$}}
\put(40,10){\framebox(10,10){$H$}}
\put(10,25){\framebox(10,10){$H$}}
\put(40,25){\framebox(10,10){$H$}}

\end{picture}

\caption{\small Quantum circuit for Deutsch's problem using one query.}
\label{fig9}
\end{figure}
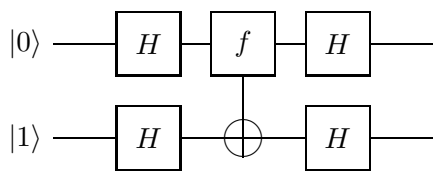

But the quantum circuit in Fig.~\ref{fig9} \cite{CEMM98,Deu85} 
computes $c_1$ with a single $f$-query gate.

\ni Here the initial state of the two-qubit system is $\ket{0}\ket{1}$ 
and its final state is $(-1)^{c_0}\ket{c_1}\ket{1}$, which yields 
$c_1$ when the first qubit is measured.

Query complexity can be pinned down more precisely than computational 
complexity in that the ``number of $f$-queries'' is not sensitive 
to arbitrary technical conventions.
So, it makes sense to consider the exact query complexity of a problem 
independent of linear factors, and to say that the classical query 
complexity of Deutsch's problem is two, whereas its quantum query 
complexity is one.

Although the above advantage is small, there are generalizations of 
Deutsch's problem for which the discrepancy between classical and 
quantum query complexity is much larger.
One of these is 
\problem{Simon's problem} \cite{Sim97}, 
which is defined as follows.
For a function $f : \01^n \rightarrow \01^n$, define $s \in \01^n$ to 
be an {\em XOR-mask\/} of $f$ if: $f(x) = f(y)$ if and only if 
$x \xor y \in \{0^n,s\}$ (where $\xor$ is defined over 
$\01^n \times \01^n$ bitwise).
When $s = 0^n$, $f$ is a bijection, and when $s \neq 0^n$, 
$f$ is a two-to-one function with a special structure related to $s$.
In Simon's problem, $f : \01^n \rightarrow \01^n$ is promised to 
have an XOR-mask $s \in \01^n$, and the goal is to find $s$ by 
making queries to $f$.
In this case, an $f$-query is the mapping $(x,y) \mapsto (x, y \xor f(x))$ 
in the classical case and $\ket{x}\ket{y} \mapsto \ket{x}\ket{y \xor f(x)}$ 
in the quantum case ($x, y \in \01^n$).
Note that Deutsch's problem is the special case of Simon's problem 
where $n = 1$ (the XOR-mask is $\neg c_1$ in this case).

It can be proven that any classical algorithm in the query model for 
Simon's problem must make $\Omega(\sqrt{2^n \log(1 / \e)})$ queries to 
$f$, even for probabilistic circuits with query gates that are permitted 
to err with probability up to $\e$.
On the other hand, there is a simple quantum circuit that solves this 
problem with only $O(n \log(1 / \e))$ queries to $f$ (see \cite{Sim97} 
for the details).
There is also a refinement to Simon's original algorithm that makes a 
polynomial number of queries and solves Simon's problem {\em exactly} 
\cite{BH97}.

Although the primary resource under consideration is the number of 
queries, the number of auxiliary operations (i.e.\ the non-query gates) 
is also of interest, and it is desirable to bound both quantities.
For Simon's algorithm the total number of gates is $O(n^2\log(1 / \e))$.

Simon's problem demonstrates that, in the query complexity setting, 
there are quantum algorithms that are exponentially more efficient 
than any classical algorithm.
Although the query complexity scenario is somewhat abstract, 
the significance of algorithms in this model will become clear 
when we examine the consequences of the 
\problem{order-finding problem in the query scenario}, 
which is defined as follows.
Let $N$ be an $n$-bit integer and $a \in \{1,\ldots,N-1\}$ be a 
number such that $\gcd(a,N) = 1$.
In this version of the order-finding problem, the function 
$f_{a,N} : \01^n \x \01^n \ra \01^n \x \01^n$ is defined as 
\begin{eqnarray}
\label{order}
f_{a,N}(x,y) = (x, (a^x y) \bmod N).
\end{eqnarray}
This is invertible if $y$ is restricted to $\{0,\ldots,N-1\}$ 
(and can be extended to be invertible over its full domain 
by defining $f_{a,N}(x,y) = (x,y)$ for the case where $N \le y < 2^n$).
The goal is to find the minimum $r \in \{1,\ldots,N-1\}$ such 
that $a^r \bmod N = 1$ by making queries to $f_{a,N}$ 
(in this case, $f_{a,N}$ is already in reversible form).
Although there is no polynomially-bounded classical circuit that 
solves this problem, Shor \cite{Sho94} discovered a quantum circuit 
that solves it with probability $1 - \e$ using only $O(\log(1 / \e))$ 
queries to $f_{a,N}$ and $O(n^2\log^d(n / \e))$ auxiliary gates 
(for some constant $d$).
Detailed explanations of the algorithm can be found in several 
sources, including \cite{CEMM98,Kit95,Sho94}.

A significant property of the function $f_{a,N}$ is that there exists 
a classical circuit of size $O(n^2\log n \log\log n)$ that takes 
$N$ (an $n$-bit number), $a \in \{1,\ldots,N-1\}$ (such that 
$\gcd(a,N)=1$), and $x, y \in \01^n$ as input, and produces 
$f_{a,N}(x,y)$ as output.
In other words, given $a$ and $N$, one can efficiently {\em simulate\/} 
an $f_{a,N}$-query gate.
Moreover, this simulation can be implemented in terms of quantum gates, 
such as {\sc not}, {\sc c-not}, and {\sc c$^2$-not} (using techniques 
for reversible classical computation \cite{Ben73}).
By doing this simulation for each $f_{a,N}$-query gate in the quantum 
circuit for the order-finding problem, one obtains a quantum circuit of 
size $O(n^2\log^d(n / \e))$ that takes $a$ and $N$ 
as input and produces the minimum positive $r$ such that $a^r \bmod N = 1$ 
as output with probability $1 - \e$.
Thus, the algorithm in the query complexity model yields an algorithm 
in the computational complexity model for order-finding---and hence 
also for factoring.
This is a specific instance of the following general result relating 
algorithms in the query complexity model to algorithms in the 
computational complexity model.

\begin{theorem}
\label{query_comp}
Suppose that a function $f_z : \01^m \ra \01^k$ is associated with 
each $z \in \01^n$ (where $m$ and $k$ are functions of $z$), 
and that the classical computational complexity of the function that 
maps $(z,x)$ to $f_z(x)$ is bounded above by $R(n)$.
Suppose also that there is a problem in the query complexity model 
where some property $P(f_z)$ is to be determined in terms of 
$f_z$-queries, and that there is a quantum circuit that solves this 
problem using $S(n)$ queries to $f_z$ and $T(n)$ auxiliary operations.
Then the quantum computational complexity of the problem where the input 
is $z \in \01^n$ and the output is the value of the property $P(f_z)$ 
is $O(R(n)S(n)+T(n))$.
\end{theorem}

The circuit for the computational complexity problem is merely the 
circuit for the query complexity problem with a circuit simulating 
each $f_z$-query gate substituted for that $f_z$-query gate.

A simple problem that seems natural in the query scenario is the 
\problem{search problem} \cite{Gro96}, 
where $f : \01^n \ra \01$, and the goal is 
to find an $x \in \01^n$ such that $f(x) = 1$ (or to indicate that 
no such $x$ exists).
Any classical algorithm for this problem must make $\Omega(2^n)$ 
$f$-queries, even if it is allowed to err with probability (say) 
$1 \over 3$.
Lov Grover \cite{Gro96} discovered a remarkable quantum algorithm that 
accomplishes this with $O(\sqrt{2^n})$ queries (some detailed 
explanations of the algorithm are found in \cite{BBHT98,Gro96,Mos98}).
Grover's result, with some later refinements 
\cite{BBHT98,BHT98,BW98,Mos98,Zal99} 
incorporated into it, is summarized as follows.

\begin{theorem}[\cite{Gro96}]
\label{grover}
There is a quantum algorithm that solves the search problem for 
$f : \01^n \ra \01$ with $O(\sqrt{2^n\log(1 / \e)})$ queries to $f$, 
and errs with probability at most $\e$.
\end{theorem}

The efficiency of the above algorithm has been shown to be optimal 
\cite{BBBV97,BBHT98,BW98,Zal97}.

Clearly, Grover's algorithm can also be used to solve the 
\problem{existential search problem}, where the goal is just to determine 
whether or not there {\em exists\/} an $x \in \01^n$ such that $f(x) = 1$ 
(a problem that also requires $\Omega(2^n)$ queries in the classical case).
Note the similarity between this existential search problem and 
the circuit satisfiability problem.
In fact, using Theorem~\ref{query_comp}, this algorithm in the query 
model translates into the algorithm for the circuit satisfiability 
problem that is claimed in Theorem~\ref{sat}.
The input is $e(C)$, an encoding of a circuit $C$ with $m$ gates and 
$n$ inputs that computes a mapping $C : \01^n \ra \01$, and the 
output should be 1 if there exists an $x \in \01^n$ such that 
$C(x)=1$, and 0 otherwise.
The mapping that takes $(e(C), x)$ to $C(x)$ can be 
computed by a classical circuit with $O(m^d)$ gates (where $d$ is 
a constant that depends on the encoding scheme, and is usually 
small).
Also, the algorithm in Theorem~\ref{grover} makes 
$O(\sqrt{2^n \log(1 / \e)}\,n)$ auxiliary operations.
Therefore, applying Theorem~\ref{query_comp}, one obtains a 
quantum circuit of size $O(\sqrt{2^n\log(1 / \e)}\,m^d)$ for the circuit 
satisfiability problem.

Let us now consider some variations and extensions of the existential 
search problem in the query model.
We shall henceforth refer to the existential search problem as {\it OR\/}, 
defined as 
\begin{eqnarray}
\mbox{\it OR\/}(f) & = & (\exists x) f(x), 
\end{eqnarray}
where $f : \01^n \ra \01$ is accessed through $f$-queries.
The name {\it OR\/} seems natural since 
\begin{eqnarray}
\mbox{\it OR\/}(f) & = & 
f(00 \cdots 0) \vee f(00 \cdots 1) \vee \cdots \vee f(11 \cdots 1).
\end{eqnarray}
Note that the complementary problem $\mbox{\it AND\/}(f) = (\forall x) f(x)$ 
has computational complexity somewhat similar to that of {\it OR}, since 
$(\forall x) f(x) = \neg (\exists x) \neg f(x)$.

Some non-trivial extensions of {\it OR\/} and {\it AND\/} are the 
\problem{alternating quantifier problems}, such as {\it OR-AND\/}, 
where there are two alternating quantifiers:
\begin{eqnarray}
\mbox{\it OR-AND\/}(f) & = & (\exists x_1)(\forall x_2) f(x_1,x_2).
\end{eqnarray}
Here, $f : \01^{n_1} \x \01^{n_2} \ra \01$, and $n_1, n_2$ are implicit 
parameters satisfying $n_1 + n_2 = n$.
By a suitable recursive application of Grover's algorithm for {\it OR\/}, 
this problem can be solved with $O(\sqrt{2^n\,n\log(1 / \e)})$ 
queries to $f$ \cite{BCW98} (the extra factor of $\sqrt{n}$ is to amplify the 
accuracy of the bottom level algorithm for {\it AND}).
In fact, one can extend the above to $k$ alternations of quantifiers: 
\begin{eqnarray}
\mbox{\it OR-AND-}\cdots\mbox{\it -Q\/}(f) & = & 
(\exists x_1)(\forall x_2) \cdots (\mbox{\sf Q} x_k) f(x_1,x_2,\ldots,x_k),
\end{eqnarray}
where $\mbox{\it Q\/} \in \{\mbox{\it OR\/},\mbox{\it AND\/}\}$ and 
$\mbox{\sf Q} \in \{\exists,\forall\}$, depending on whether 
$k$ is even or odd, and $f : \01^{n_1} \x \cdots \x \01^{n_k} \ra \01$ 
with $n_1 + \cdots n_k = n$.
In this case, the recursive application of Grover's technique makes 
$O(\sqrt{2^n\,n^{k-1}\log(1 / \e)})$ queries to $f$ (see \cite{BCW98}; 
also \cite{Ozh98} for a related result).

For all of these variations of {\it OR\/} and {\it AND}, it can be shown 
that any classical algorithm for one of these problems must make 
$\Omega(2^n)$ queries, and the quantum algorithms for these 
problems are all nearly quadratically more efficient than this in 
the sense that they make $O((2^n)^{1/2+\delta})$ queries, for any 
$\delta > 0$ and $\e > 0$.
In fact, even if $k$, the number of alternations of {\it OR\/} and 
{\it AND}, is set to $\delta n / 2 \log n$ (instead of being held 
constant), the quantum algorithms make $O((2^n)^{1/2+\delta})$ queries.
All of these quantum algorithms also have counterparts for the 
corresponding problems in the computational model, where the 
function is specified by an encoding $e(C)$ of a circuit $C$.

Another problem that has a similar flavor to these problems is the 
\problem{parity problem (in the query scenario)}, defined as 
\begin{eqnarray}
\label{parity}
\mbox{\it PARITY\/}(f) & = & 
\left(\sum_{x \in \01^m} f(x) \right) \bmod 2.
\end{eqnarray}
It can be shown that any classical algorithm requires $\Omega(2^n)$ 
queries to solve {\it PARITY\/}, and it is natural to ask whether 
quantum algorithms can be quadratically more efficient---or even 
$O((2^n)^r)$, for some $r < 1$.
One of the applications of the communication complexity model 
(explained in the next section) is to show that this is not possible: 
at least $\Omega(2^n/n)$ queries must be made by any quantum algorithm.
In fact, a stronger lower bound of $\half 2^n$ is also known 
\cite{BBCMW98,FGGS98} (using different methods).

It is important to note that, although upper bounds in the query model 
translate into upper bounds in the computational model, the converse 
of this need not be true.
For example, it is conceivable that there is a polynomially-bounded 
circuit that solves the 
\problem{circuit parity problem}, 
where the input is $e(C)$, an encoding of a circuit $C$ 
that computes a function $f$, and the output is ${\it PARITY\/}(f)$.
Note how this latter problem is different from another version of the 
parity problem in the computational scenario (discussed in Section 1.1), 
where the inputs are $x_0,x_1,\ldots,x_{n-1}$, and the goal is to 
compute $x_0 \xor x_1 \xor \cdots \xor x_{n-1}$.

\section{Communication complexity}

In this model, there are two parties, traditionally referred 
to as Alice and Bob, who each receive an $n$-bit binary 
string as input ($x = x_0 x_1 \ldots x_{n-1}$ for Alice and 
$y = y_0 y_1 \ldots y_{n-1}$ for Bob) and the goal is for 
them to determine the value of some function of the 
of these $2n$ bits.
The resource under consideration here is the {\em communication\/} 
between the two parties, and an algorithm is a {\em protocol}, 
where the parties send information to each other (possibly in both 
directions and over several rounds) until one of them (say, Bob) 
obtains the answer.
This model was introduced by Yao \cite{Yao79} and has been widely 
studied in the classical context (see \cite{KN98} for a survey).

An interesting example is the 
\problem{equality problem}, 
where the function is {\it EQ\/}, defined as 
\begin{eqnarray*}
{\it EQ\/}(x,y) & = & \cases{ 1 & if $x=y$\cr
                             0 & if $x \neq y$.\cr}
\end{eqnarray*}
A simple $n$-bit protocol for {\it EQ\/} is for Alice to just send 
her bits $x_0,\ldots,x_{n-1}$ to Bob, after which Bob can evaluate the 
function by himself (in fact, there is a similar $n$-bit protocol for 
{\em any\/} function).
The interesting question is whether or not the {\it EQ\/} function 
can be evaluated with fewer than $n$ bits of communication---after 
all, the goal here is only for Bob to acquire one bit.
The answer depends on whether or not any error probability is 
permitted.

If Bob must acquire the value of $\mbox{\it EQ\/}(x,y)$ with certainty 
then it turns out that $n$ bits of communication are necessary.
Note that Alice sending the first $n-1$ bits of $x$ will clearly 
{\em not\/} work, since the answer could critically depend on 
whether or not $x_{n-1} = y_{n-1}$.
The number of possible protocols to consider is quite large and an 
actual {\em proof\/} that $n$ bits communication are necessary is 
nontrivial.
The interested reader is referred to \cite{KN98} for a proof.

On the other hand, for probabilistic protocols (where Alice and Bob 
can flip coins and base their behavior on the outcomes), if an error 
probability of $\e > 0$ is permitted then $O(\log(n)\log(1/\e))$ 
bits of communication are sufficient.
As usual, we are not assuming anything about a probability distribution 
on the input strings; the error probability is with respect to the 
random choices made by Alice and Bob, and it applies regardless of 
what $x$ and $y$ are.

We now describe an $O(\log(n)\log(1/\e))$-bit protocol for {\it EQ}.
First of all, Alice and Bob agree on a finite field whose size is 
between $2n$ and $4n$ (such a field always exists, and its elements 
can be represented as $O(\log(n))$-bit strings).
Now, consider the two polynomials 
\begin{eqnarray}
p_x(t) & = & x_0 + x_1 t + \cdots + x_{n-1} t^{n-1} \\
p_y(t) & = & y_0 + y_1 t + \cdots + y_{n-1} t^{n-1}.
\end{eqnarray}
For any value of $t$ in the field, Alice can evaluate $p_x(t)$ and 
Bob can evaluate $p_y(t)$.
If $x=y$ then the two polynomials are identical, so $p_x(t) = p_y(t)$ 
for every value of $t$.
But, if $x \neq y$ then, since $p_x(t)$ and $p_y(t)$ are polynomials of 
degree $n-1$, there can be at most $n-1$ distinct values of $t$ for 
which $p_x(t) = p_y(t)$.
Therefore, if a value of $t$ is chosen randomly from the field then 
the probability that $p_x(t) = p_y(t)$ is at most $\half$.
Now, the protocol proceeds as follows.
Alice chooses $k = \log(1 / \e)$ independent random elements of 
the field, $t_1,\ldots,t_k$, and then sends $t_1,\ldots,t_k$ and 
$p_x(t_1),\ldots,p_x(t_k)$ to Bob (this consists of $O(\log(n)\log(1/\e))$ 
bits).
Then Bob outputs 1 if and only if 
$p_x(t_i) = p_y(t_i)$ for all $i \in \{1,\ldots,k\}$.
The probability that Bob erroneously outputs 1 when $x \neq y$ 
is at most $1/2^k = \e$.

Two other interesting communication complexity problems are the 
\problem{intersection problem}, 
where the function is {\it IN\/}, defined as 
\begin{eqnarray}
\mbox{\it IN\/}(x,y) 
%& = & 
%\cases{ 1 & if $x_i \wedge y_i = 1$ for some $i \in \{0,\ldots,n-1\}$\cr
%        0 & otherwise\cr} \nonumber \\
& = & (x_0 \wedge y_0) \vee (x_1 \wedge y_1) \vee \cdots 
\vee (x_{n-1} \wedge y_{n-1})
\end{eqnarray}
and the \problem{inner product problem}, 
where the function is {\it IP\/}, defined as 
\begin{eqnarray}
\mbox{\it IP\/}(x,y) 
%& = & (x_0 y_0 + \cdots + x_{n-1} y_{n-1}) \bmod 2 \nonumber \\
& = & (x_0 \wedge y_0) \xor (x_1 \wedge y_1) \xor \cdots 
\xor (x_{n-1} \wedge y_{n-1}).
\end{eqnarray}
Intuitively, for {\it IN}, the inputs $x$ and $y$ can be thought 
as encodings of two subsets of $\{0,\ldots,n-1\}$ and the output 
is a bit indicating whether or not they intersect.
Also, {\it IP\/} is the inner product of $x$ and $y$ as bit vectors 
in modulo two arithmetic.
The deterministic communication complexity of each of these problems 
is the same as that of {\it EQ\/}: any deterministic protocol requires 
$n$ bits of communication.
Also, it has been shown that both of these problems are more difficult 
than {\it EQ\/} when probabilistic protocols are considered: any 
probabilistic protocol with error probability up to (say) $1 \over 3$ 
requires $\Omega(n)$ bits of communication (see \cite{CG88} for {\it IP}, 
and \cite{KS87} for {\it IN\/}; also \cite{KN98}).

It is natural to ask whether any reduction in communication can be 
obtained by somehow using {\em quantum\/} information.
Define a {\em quantum\/} communication protocol as one where Alice 
and Bob can exchange messages that consist of qubits.
In a more formal definition of this model, there is an {\it a priori\/} 
system of $m$ qubits, some of them in Alice's possession and some of 
them in Bob's possession.
The initial state of all of these qubits can be assumed to be $\ket{0}$, 
and Alice and Bob can each perform unitary transformations on those 
qubits that are in their possession and they can also send qubits between 
themselves (thereby changing the ownership of qubits).
The output is then taken as the outcome of some measurement of Bob's 
qubits.
Various preliminary results about communication complexity with 
quantum information occurred in \cite{BCD97,CB97,DHT97,Kre95,Yao93}.

There are fundamental results in quantum information theory which 
imply that classical information cannot be ``compressed'' within 
quantum information \cite{Hol73}.
For example, Alice cannot convey more than $r$ classical bits of 
information to Bob by sending him an $r$-qubit message.
Based on this, one might mistakenly think that there is no 
advantage to using quantum information in the communication 
complexity context.
In fact, there exists a quantum communication protocol that 
solves {\it IN\/} whose qubit communication is approximately the 
square root of the bit communication of the best possible 
classical probabilistic protocol.

\begin{theorem}[\cite{BCW98}]
\label{intersect}
There exists a quantum protocol for the intersection problem ({\it IN}) 
that uses $O(\sqrt{n\log(1 / \e)}\log(n))$ qubits of communication and errs 
with probability at most $\e$.
\end{theorem}

\ni Moreover, the quantum protocol can be adapted to actually find 
a point in the intersection in the cases where $\mbox{\it IN\/}(x,y) = 1$.
That is, to produce an $i \in \{0,\ldots,n-1\}$ such that 
$x_i \wedge y_i = 1$.
This problem, like {\it IN}, has classical probabilistic communication 
complexity $\Omega(n)$.

To understand the protocol in Theorem~\ref{intersect}, it is helpful 
to think of the inputs $x$ and $y$ as functions rather than strings, 
and we introduce some notation that makes this explicit.
For convenience, assume that $n = 2^k$ for some $k$ 
(if not then $x$ and $y$ can lengthened by padding them 
with zeroes), and define the functions $f_x, f_y : \01^k \ra \01$ as 
\begin{eqnarray}
f_x(i) & = & x_i \\
f_y(i) & = & y_i
\end{eqnarray}
where $\01^k$ and $\{0,1,\ldots,2^k-1\}$ are identified in the 
natural way.
Alice and Bob's input data can be thought of as $f_x$ and $f_y$, rather 
than $x$ and $y$ (respectively).
In particular, given $x$, Alice can simulate an $f_x$-query 
that maps $\ket{i}\ket{j}$ to $\ket{i}\ket{j \xor f_x(i)}$ 
(for all $i \in \01^k$ and $j \in \01$), and Bob can simulate 
$f_y$-queries.
(Although the resource that is of interest in this model is not the 
number of basic operations that Alice and Bob perform, it is worth 
noting that, Alice and Bob's simulations of these queries can be 
explicitly implemented by reversible circuits with 
$O(2^k k) = O(n \log(n))$ basic operations).

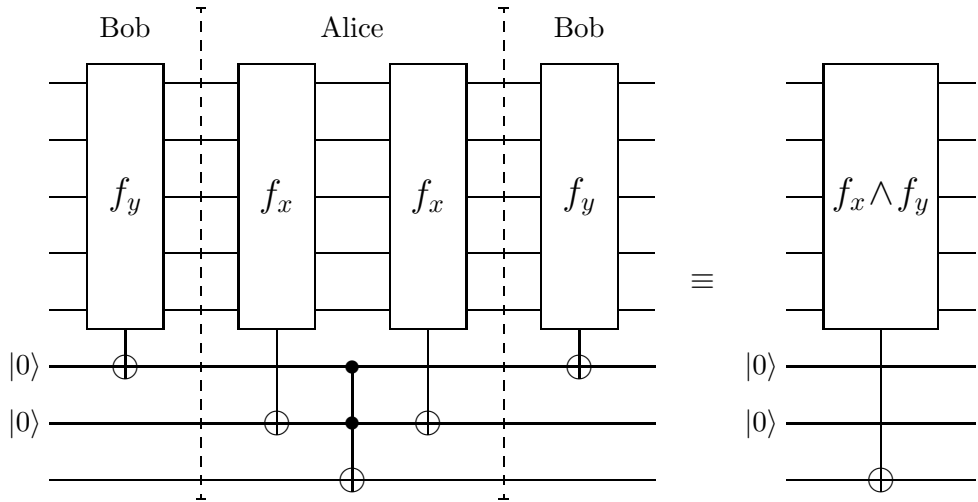
\begin{figure}[h]
\centering

\setlength{\unitlength}{0.0198in}

\begin{picture}(7,135)(0,0)
\put(3,35){\makebox(0,0){$\ket{0}$}}
\put(3,20){\makebox(0,0){$\ket{0}$}}
\end{picture}
\begin{picture}(160,135)(0,0)

\put(0,5){\line(1,0){160}}
\put(0,20){\line(1,0){160}}
\put(0,35){\line(1,0){160}}

\put(0,50){\line(1,0){10}}
\put(0,65){\line(1,0){10}}
\put(0,80){\line(1,0){10}}
\put(0,95){\line(1,0){10}}
\put(0,110){\line(1,0){10}}

\put(30,50){\line(1,0){20}}
\put(30,65){\line(1,0){20}}
\put(30,80){\line(1,0){20}}
\put(30,95){\line(1,0){20}}
\put(30,110){\line(1,0){20}}

\put(70,50){\line(1,0){20}}
\put(70,65){\line(1,0){20}}
\put(70,80){\line(1,0){20}}
\put(70,95){\line(1,0){20}}
\put(70,110){\line(1,0){20}}

\put(110,50){\line(1,0){20}}
\put(110,65){\line(1,0){20}}
\put(110,80){\line(1,0){20}}
\put(110,95){\line(1,0){20}}
\put(110,110){\line(1,0){20}}

\put(150,50){\line(1,0){10}}
\put(150,65){\line(1,0){10}}
\put(150,80){\line(1,0){10}}
\put(150,95){\line(1,0){10}}
\put(150,110){\line(1,0){10}}

\put(10,45){\framebox(20,70){\Large $f_y$}}
\put(50,45){\framebox(20,70){\Large $f_x$}}
\put(90,45){\framebox(20,70){\Large $f_x$}}
\put(130,45){\framebox(20,70){\Large $f_y$}}

\put(20,125){\makebox(0,0){Bob}}
\put(80,125){\makebox(0,0){Alice}}
\put(140,125){\makebox(0,0){Bob}}

\put(20,32){\line(0,1){13}}
\put(60,17){\line(0,1){28}}
\put(100,17){\line(0,1){28}}
\put(140,32){\line(0,1){13}}
\put(20,35){\circle{6}}
\put(60,20){\circle{6}}
\put(100,20){\circle{6}}
\put(140,35){\circle{6}}
\put(80,2){\line(0,1){33}}
\put(80,5){\circle{6}}
\put(80,20){\circle*{3.4}}
\put(80,35){\circle*{3.4}}

\thinlines
\put(40,0){\dashbox{2.5}(0,130){}}
\put(120,0){\dashbox{2.5}(0,130){}}
%\put(40,0){\line(0,1){125}}
%\put(120,0){\line(0,1){125}}

\end{picture}
\begin{picture}(20,135)(0,0)

\put(0,0){\makebox(20,115){\large $\equiv$}}

\end{picture}
\begin{picture}(7,135)(0,0)
\put(3,35){\makebox(0,0){$\ket{0}$}}
\put(3,20){\makebox(0,0){$\ket{0}$}}
\end{picture}
\begin{picture}(50,135)(0,0)

\put(0,5){\line(1,0){50}}
\put(0,20){\line(1,0){50}}
\put(0,35){\line(1,0){50}}

\put(0,50){\line(1,0){10}}
\put(0,65){\line(1,0){10}}
\put(0,80){\line(1,0){10}}
\put(0,95){\line(1,0){10}}
\put(0,110){\line(1,0){10}}

\put(40,50){\line(1,0){10}}
\put(40,65){\line(1,0){10}}
\put(40,80){\line(1,0){10}}
\put(40,95){\line(1,0){10}}
\put(40,110){\line(1,0){10}}

\put(10,45){\framebox(30,70){\Large $f_x\!\wedge\!f_y$}}

\put(25,2){\line(0,1){43}}
\put(25,5){\circle{6}}

\end{picture}
\caption{\small Simulation of an $(f_x \wedge f_y)$-query in terms 
of $f_x$-queries and $f_y$-queries.}
\label{fig10}
\end{figure}

To construct an efficient quantum protocol for {\it IN}, define the 
function $f_x \wedge f_y : \01^k \ra \01$ as 
$(f_x \wedge f_y)(i) = f_x(i) \wedge f_y(i)$ (for $i \in \01^k$), and 
note that $\mbox{\it IN\/}(x,y) = \mbox{\it OR\/}(f_x \wedge f_y)$.
Therefore, if Alice and Bob can somehow perform $(f_x \wedge f_y)$-queries
then the value of $\mbox{\it IN\/}(x,y)$ can be determined by making 
$O(\sqrt{2^k\log(1 / \e)}) = O(\sqrt{n\log(1 / \e)})$ such queries.
The problem is that neither Alice nor Bob individually have enough 
information to perform an $(f_x \wedge f_y)$-query (since this 
depends on both $x$ and $y$).
If Alice were to begin by sending $x$ to Bob then Bob could make 
$(f_x \wedge f_y)$-queries on his own, but note that this entails 
$n$ bits of communication to begin with.
Another, more efficient, approach is for Alice and Bob to collectively 
simulate $(f_x \wedge f_y)$-queries by combining $f_x$-queries (which 
Alice can perform) with $f_y$-queries (which Bob can perform), and 
a small amount of communication.
To see how this is accomplished, consider the circuit in Fig.~\ref{fig10}.
First, ignoring the broken vertical lines, note that the quantum 
circuit (composed of two $f_x$-queries, two $f_y$-queries, and one 
Toffoli gate) is equivalent to an $(f_x \wedge f_y)$-query.
That is, it implements the unitary transformation that maps the state 
$\ket{i}\ket{0}\ket{0}\ket{j}$ to the state 
$\ket{i}\ket{0}\ket{0}\ket{j \xor (f_x \wedge f_y)(i)}$ 
(for all $i \in \01^k$, $j \in \01$).
This circuit uses two extra qubits that are each initialized 
in state $\ket{0}$ and which incur no net change.

Now, the protocol for {\it IN\/} can be thought of as Bob executing 
the algorithm in the query model for {\it OR\/} with the function 
$f_x \wedge f_y$, except that, whenever an $(f_x \wedge f_y)$-query 
gate arises, he interacts with Alice to simulate the circuit in 
Fig.~\ref{fig10}: first Bob performs an $f_y$-query gate, then he 
sends the $k + 3$ qubits to Alice who performs some actions involving 
$f_x$-queries and a Toffoli gate (shown between the two broken 
lines) and sends the qubits back to Bob, who performs another $f_y$-query.
Note that the total amount of communication that this entails is 
$2(k+3) \in O(\log n)$ qubits.
Therefore, the total communication for Bob's simulation of the 
$O(\sqrt{n \log (1 / \e)})$ queries to $(f_x \wedge f_y)$ is 
$O(\sqrt{n \log (1 / \e)}\log(n))$, as claimed in Theorem~\ref{intersect}.

More recently, Ran Raz has given an example of a communication 
complexity problem which a quantum protocol can solve with 
{\em exponentially\/} less communication than the best classical 
probabilistic protocol.
The description of the problem is more complicated than {\it EQ}, 
{\it IN}, and {\it IP}, and the reader is referred to \cite{Raz99} for 
the details.

The methodology used to establish Theorem~\ref{intersect} 
involved the conversion of an algorithm in the query model 
(for {\it OR\/}) to a communication protocol 
(for $\mbox{\it IN\/}(x,y) = \mbox{\it OR\/}(f_x \wedge f_y)$).
This conversion can be stated in a more general form.

\begin{theorem}[\cite{BCW98}]
\label{query_comm}
Suppose that there is a quantum algorithm in the query model that 
computes $\mbox{\it P\/}(f)$ in terms of\, $T(k,\e)$ queries to $f$, 
where $f : \01^k \ra \01$, and $\e$ is a bound on the error 
probability.
For $n = 2^k$, define the communication problem 
$P_{\mbox{\tiny $\wedge$}} : \01^n \x \01^n \ra \01$ 
as $P_{\mbox{\tiny $\wedge$}}(x,y) = P(f_x \wedge f_y)$.
Then there is a quantum protocol that solves $P_{\mbox{\tiny $\wedge$}}$ 
with $O(T(\log(n),\e)\log(n))$ qubits of communication.
And a similar result holds for 
$P_{\mbox{\tiny $\vee$}}(x,y) = P(f_x \vee f_y)$ and 
$P_{\mbox{\tiny $\xor$}}(x,y) = P(f_x \xor f_y)$.
\end{theorem}

We conclude with a discussion of the quantum communication complexity 
of the inner product function {\it IP}.
It has been shown \cite{Kre95} (see also \cite{CDNT97}) that even quantum 
protocols require communication $\Omega(n)$ for this problem, even 
when the error probability is permitted to be as large as (say) 
$1 \over 3$.
This fact, combined with Theorem~\ref{query_comm} applied in its 
contrapositive form, can be used to establish a lower bound for 
the parity problem in the query model (defined in Eq.~\ref{parity}).
The main observation is that 
$\mbox{\it IP\/}(x,y) = {\it PARITY\/}(f_x \wedge f_y)$.
Suppose that there is a quantum algorithm that computes  
$\mbox{\it PARITY\/}(f)$ for $f : \01^k \ra \01$ by making 
$T(k)$ $f$-queries (assume that the error probability is bounded 
by $1 \over 3$).
Then, by Theorem~\ref{query_comm}, there exists a quantum protocol 
that solves {\it IP\/} with $O(T(k) k)$ qubits 
of communication, where $n = 2^k$ is the size of the input instance 
to {\it IP}.
Since there is a lower bound of $\Omega(n) = \Omega(2^k)$ for the 
communication complexity of {\it IP}, we must have 
$T(k) k \in \Omega(2^k)$, which implies that 
$T(k) \in \Omega(2^k / k)$.
This is an easy way to get a ``ball park'' lower bound for the query 
complexity of {\it PARITY}, whose exact value is known to be 
$\half 2^k$ by other methods \cite{BBCMW98,FGGS98}.

\section*{Acknowledgments}

I would like to thank Michele Mosca for comments and help with 
the references.
This research was partially supported by the Natural Sciences and 
Engineering Research Council of Canada.


\begin{thebibliography}{9}

\bibitem{AB96}
D.~Aharonov and M.~Ben-Or, 
\newblock ``Fault-tolerant quantum computation with constant error'',
\newblock {\em Proc.\ 29th Ann.\ ACM Symp.\ on Theory of Computing 
(STOC '97)}, pp.~176--188, 1997.

\bibitem{APR83}
L.~Adleman, C.~Pomerance, R.~Rumely, 
\newblock ``On distinguishing prime numbers from composite numbers'', 
\newblock {\em Annals of Mathematics}, Vol.~117, pp.~173--206, 1983.

\bibitem{BBC95}
A.~Barenco, C.H.~Bennett, R.~Cleve, D.P.~DiVincenzo, N.~Margolus, 
P.~Shor, T.~Sleator, J.~Smolin, and H.~Weinfurter, 
\newblock ``Elementary gates for quantum computation'',
\newblock {\em Phys.\ Rev.\ A}, Vol.~52, pp.~3457--3467, 1995.

\bibitem{BBCMW98}
R.~Beals, H.~Buhrman, R.~Cleve, M.~Mosca, R.~de~Wolf, 
\newblock ``Quantum lower bounds by polynomials'', 
\newblock {\em Proc.\ 39th Ann.\ IEEE Symp.\ on Foundations of Computer 
Science (FOCS '98)}, pp.~352--361, 1998.

\bibitem{Ben73}
C.H.~Bennett, 
\newblock ``Logical reversibility of computation'', 
\newblock {\em IBM J. of Research and Development}, Vol.~17, 
pp.~525--532, 1973.

\bibitem{BBBV97}
C.H.~Bennett, E.~Bernstein, G.~Brassard, and U.~Vazirani,
\newblock ``Strengths and weaknesses of quantum computing'', 
\newblock {\em SIAM J. on Computing}, 
Vol.~26, No.~5, pp.~1510--1523, 1997.

\bibitem{BV93}
E.~Bernstein and U.V.~Vazirani, 
\newblock ``Quantum complexity theory'', 
\newblock {\em SIAM J. on Comput.}, 
Vol.~26, No.~5, pp.~1411--1473, 1997.

\bibitem{BBHT98}
M.~Boyer, G.~Brassard, P.~H{\o}yer, and A.~Tapp, 
``Tight bounds on quantum searching'', 
\newblock {\em Fortschritte der Physik}, Vol.~46, pp.~493--505, 1998.
\newblock (An earlier version appeared in {\em Physcomp '96}.)

\bibitem{BMPRV99}
P.O.~Boykin, T.~Mor, M.~Pulver, V.~Roychowdhury, and F.~Vatan, 
``On universal and fault-tolerant quantum computing'', 
\newblock preprint quant-ph/9906054, 1999.

\bibitem{BHT98}
G.~Brassard, P.~H{\o}yer, and A.~Tapp, 
\newblock ``Quantum counting'', 
\newblock {\em Proc.\ 25th ICALP}, Vol.~1443 of {\em Lecture Notes
  in Computer Science}, pp.~820--831 (Springer), 1998.

\bibitem{BH97}
G.~Brassard and P.~H{\o}yer, 
\newblock ``An exact quantum polynomial-time algorithm for {S}imon's 
problem'',
\newblock {\em Proc.\ 5th Israeli Symp.\ on Theory of Computing and 
Systems (ISTCS '97)}, pp.~12--23, 1997.

\bibitem{BCD97}
H.~Buhrman, R.~Cleve, and  W.~van Dam, 
\newblock ``Quantum entanglement and communication complexity'',
\newblock preprint quant-ph/9705033, 1997.

\bibitem{BCW98}
H.~Buhrman, R.~Cleve, and A.~Wigderson, 
\newblock ``Quantum vs. classical communication and computation'',
\newblock {\em Proc.\ 30th Ann.\ ACM Symp.\ on Theory of Computing 
(STOC '98)}, pp.~63-68, 1998.

\bibitem{BW98}
H.~Buhrman, R.~de~Wolf, 
\newblock ``Lower bounds for quantum search and derandomization'', 
\newblock preprint quant-ph/9811046, 1998.

\bibitem{CG88}
B.~Chor and O.~Goldreich, 
\newblock ``Unbiased bits from sources of weak randomness and probabilistic 
communication complexity'', 
\newblock {\em SIAM J. Comput.}, Vol.~17, No.~2, pp. 230--261, 1988.

\bibitem{CB97}
R.~Cleve and H.~Buhrman, 
\newblock ``Substituting quantum entanglement for communication'', 
\newblock {\em Phys.\ Rev.\ A}, Vol.~56, No.~2, pp.~1201--1204, 1997.

\bibitem{CDNT97}
R.~Cleve, W.~van Dam, M.~Nielsen, and A.~Tapp, 
\newblock ``Quantum entanglement and the communication complexity of the 
inner product function'', 
\newblock to appear in {\em Proc.\ 1st NASA Intl.\ Conf.\ on Quantum 
Computing and Quantum Communications}, 
Vol.~1509 of {\em Lecture Notes in Computer Science} (Springer), 1998.
\newblock preprint quant-ph/9708019, 1997.

\bibitem{CEMM98}
R.~Cleve, A.~Ekert, C.~Macchiavello, and M.~Mosca, 
\newblock ``Quantum algorithms revisited'', 
\newblock {\em Proc.\ of the Royal Society of London}, Vol.~A454, 
pp.~339--354, 1998.

\bibitem{Coo71}
S.A.~Cook, 
\newblock ``The complexity of theorem proving procedures'', 
\newblock {\em Proc.\ 3rd Ann.\ ACM Symp.\ on Theory of Computing 
(STOC '71)}, pp.~151--158, 1971.

\bibitem{DHT97}
W.~van~Dam, P.~H\o yer, A.~Tapp, 
\newblock ``Multiparty quantum communication complexity'', 
\newblock preprint quant-ph/9710054, 1997.

\bibitem{Deu85}
D.~Deutsch, 
\newblock ``Quantum theory, the Church-Turing principle and the universal 
quantum computer'', 
\newblock {\em Proc.\ of the Royal Society of London}, Vol.~A400, 
pp.~96--117, 1985.

\bibitem{Deu89}
D.~Deutsch, 
\newblock ``Quantum computational networks'', 
\newblock {\em Proc.\ of the Royal Society of London}, Vol.~A425, 
pp.~73--90, 1989.

\bibitem{Div98}
D.P.~DiVincenzo, 
\newblock ``Quantum gates and circuits'', 
\newblock {\em Proc.\ of the Royal Society of London}, Vol.~A454, 
pp.~261--276, 1998.

\bibitem{FGGS98}
E.~Farhi, J.~Goldstone, S.~Gutmann, and M.~Sipser, 
\newblock ``A limit on the speed of quantum computation in determining 
parity.''
\newblock {\em Phys.\ Rev.\ Lett.}, Vol.~81, pp.~5442--5444, 1998.

\bibitem{FR98}
L.~Fortnow and J.~Rogers, 
\newblock ``Complexity limitations on quantum computation'', 
\newblock {\em Proc.\ 13th IEEE Conf.\ on Computational Complexity}, 
pp.~202--209, 1998.

\bibitem{GJ79}
M.R.~Garey and D.S.~Johnson, 
\newblock {\em Computers and Intractibility: A Guide to the Theory of 
NP-Completeness}, W.H.~Freeman, 1979.

\bibitem{Gro96}
L.K.~Grover, 
\newblock ``A fast quantum mechanical algorithm for database search'', 
\newblock {\em Proc.\ 28th Ann.\ ACM Symp. on Theory of Computing 
(STOC '96)}, pp.~212--219, 1996.

\bibitem{Hol73}
A.S.~Holevo, 
\newblock ``Some estimates of the information transmitted by 
quantum communication channels'', 
\newblock {\em Problemy Peredachi Informatsii}, Vol.~9, pp.~3--11, 1973.
\newblock English translation in {\em Problems of Information 
Transmission (USSR)}, Vol.~9, pp.~177--183, 1973.

\bibitem{KS87}
B.~Kalyanasundaram and G.~Schnitger, 
\newblock ``The probabilistic communication complexity of set 
intersection'',
\newblock {\em Proc.\ 2nd Conf.\ on Structure in Complexity Theory}, 
pp.~41--49, 1987.

\bibitem{Kni95}
E.~Knill, 
\newblock personal communication, 1996.

\bibitem{KLZ96}
E.~Knill, R.~Laflamme, W.~Zurek, 
\newblock ``Threshold accuracy for quantum computation'', 
\newblock preprint quant-ph/9610011, 1996.

\bibitem{Kit95}
A.Y.~Kitaev, 
\newblock ``Quantum measurements and the Abelian stabilizer problem'', 
\newblock preprint quant-ph/9511026, 1995.

\bibitem{Kit97}
A.Y.~Kitaev, 
\newblock ``Quantum computations: algorithms and error correction'', 
\newblock {\em Russian Math.\ Surveys}, 
\newblock Vol.~52, No.~6, pp.~1191-1249, 1997.

\bibitem{Kre95}
I.~Kremer, 
\newblock {\em Quantum Communication}, 
\newblock MSc Thesis, Computer Science Department, The Hebrew University, 1995.

\bibitem{KN98}
E.~Kushilevitz and N.~Nisan, 
\newblock {\em Communication Complexity}, 
\newblock (Cambridge University Press), 1998.

\bibitem{LP92}
H.~Lenstra and C.~Pomerance, 
\newblock ``A rigorous time bound for factoring integers'', 
\newblock {\em J. of the AMS}, Vol.~5, No.~2, pp.~482--516, 1992.

\bibitem{Mos98}
M.~Mosca, 
\newblock ``Quantum searching, counting and amplitude amplification by
eigenvector analysis'', 
\newblock {\em MFCS '98 workshop on Randomized Algorithms}, 1998.

\bibitem{ME98}
M.~Mosca and A.~Ekert, 
\newblock ``The hidden subgroup problem and eigenvalue estimation on 
a quantum computer'', 
\newblock to appear in {\em Proc.\ 1st NASA Intl.\ Conf.\ on Quantum 
Computing and Quantum Communications}, 
Vol.~1509 of {\em Lecture Notes in Computer Science} (Springer), 1998.
\newblock preprint quant-ph/9903071, 1998.

\bibitem{NO99}
H.~Nishimura and M.~Ozawa, 
\newblock ``Computational complexity of uniform quantum circuit families 
and quantum Turing machines'',
\newblock preprint quant-ph/9906095, 1999.

\bibitem{Ozh98}
Y.~Ozhigov, 
\newblock ``Fast quantum verification for the formulas of 
predicate calculus'', 
\newblock preprint quant-ph/9809015, 1998.

\bibitem{Pom87}
C.~Pomerance, 
\newblock ``Fast rigorous factorization and discrete logarithm 
algorithms'', 
\newblock {\em Discrete Algorithms and Complexity (Proc.\ Japan-US 
Joint Seminar on Discrete Algorithms and Complexity theory)}, 
pp.~119--143, 1987.

\bibitem{Pre98}
\newblock J.~Preskill, 
\newblock ``Fault-tolerant quantum computation'', 
\newblock in {\em Introduction to Quantum Computation and Information}, 
edited by H.-K.~Lo, S.~Popescu, and T.P.~Spiller (World Scientific), 
pp.~213--269, 1998.

\bibitem{Raz99}
\newblock R.~Raz,
\newblock ``Exponential separation of quantum and classical 
communication complexity'',
\newblock {\em Proc.\ 31st Ann.\ ACM Symp.\ on Theory of 
Computing (STOC '99)}, pp.~358--367, 1999.

\bibitem{Sho94}
P.W.~Shor, 
\newblock ``Polynomial-time algorithms for prime factorization and discrete
  logarithms on a quantum computer'', 
\newblock {\em SIAM J. on Computing}, Vol.~26, No.~5, pp.~1484--1509, 1997.
\newblock (An earlier version appeared in {\em FOCS '94}.)

\bibitem{Sho97}
P.W.~Shor, 
\newblock ``Fault-tolerant quantum computation'', 
\newblock {\em Proc.\ 37th Ann.\ IEEE Symp.\ on Foundations of Computer 
Science (FOCS '96)}, pp.~56-65, 1996.

\bibitem{Sim97}
D.~Simon, 
\newblock ``On the power of quantum computation'', 
\newblock {\em SIAM J. on Computing}, Vol.~26, No.~5, pp.~1474--1483, 1997.
\newblock (An earlier version appeared in {\em FOCS '94}.)

\bibitem{SW95}
T.~Sleator and H.~Weinfurter, 
\newblock ``Realizable universal quantum logic gates'', 
\newblock {\em Phys.\ Rev.\ Lett.}, Vol.~74, pp.~4087--4090, 1995.

\bibitem{Sol99}
R.~Solovay, 
\newblock ``Lie groups and quantum circuits'', 
paper in preparation, 1999.

\bibitem{SS77}
R.~Solovay and V.~Strassen, 
\newblock ``A fast Monte Carlo test for primality'', 
\newblock {\em SIAM J. on Computing}, Vol.~6, pp.~84--85, 1977.
\newblock (Erratum in Vol.~7, p.~118, 1978.)

\bibitem{Vaz87}
U.V.~Vazirani, 
\newblock ``Strong communication complexity or generating quasi-random 
sequences from two communicating slightly-random sources'',
\newblock {\em Combinatorica}, Vol.~7, No.~4, pp.~375--392, 1987.

\bibitem{Yao79}
A.~C.-C.~Yao, 
\newblock ``Some complexity questions related to distributive computing",
\newblock {\em Proc.\ 11th Ann.\ ACM Symp.\ on Theory of Computing 
(STOC '79)}, pp.~209-213, 1979.

\bibitem{Yao93}
A.~C.-C.~Yao, 
\newblock ``Quantum circuit complexity'', 
\newblock {\em Proc.\ 34th Ann.\ IEEE Symp.\ on Foundations of 
Computer Science (FOCS '93)}, pp.~352--361, 1993.

\bibitem{Zal97}
C.~Zalka, 
\newblock ``Grover's quantum searching algorithm is optimal'', 
\newblock preprint quant-ph/9711070, 1997.

\bibitem{Zal99}
C.~Zalka, 
\newblock ``A Grover-based quantum search of optimal order for an 
unknown number of marked elements'', 
preprint quant-ph/9902049, 1999.

\end{thebibliography}
\end{document}